\documentclass[epj]{svjour}
\usepackage{graphicx}
\newlength{\colw}
\begin{document}
\title{Quark Matter in QC$_2$D}
\author{Simon Hands\inst{1}, Seyong Kim\inst{2} \and Jon-Ivar Skullerud\inst{3}
}                     % Do not remove
\institute{Department of Physics, Swansea University, Singleton Park,
Swansea SA2 8PP, U.K.\and 
Department of Physics, Sejong University, Gunja-Dong, Gwangjin-Gu, Seoul
143-747, South Korea
\and
School of Mathematics, Trinity College, Dublin 2, Ireland
}
\date{Received: date / Revised version: date}
\abstract{
Results are presented from a numerical study of lattice QCD with gauge group 
SU(2) and two flavors of Wilson fermion at non-zero quark chemical potential
$\mu\gg T$.
Studies of the equation of state, the superfluid condensate, and the Polyakov
line all suggest that in addition to the low density phase of Bose-condensed
diquark baryons, there is a deconfined phase at higher quark density in which 
quarks form a degenerate system, whose Fermi surface is only mildly 
disrupted by Cooper pair condensation.
 % end of PACS codes
} %end of abstract
\maketitle
\section{Introduction}
\label{intro}
The phase structure of QCD at large baryon density is one of the most
fascinating areas of strong interaction physics, and yet a systematic
calculational approach to this problem remains elusive. Lattice QCD 
simulation, the
usual non-perturbative approach of choice, fails dismally because in Euclidean
metric the quark action 
$\bar qM(\mu)q$, where $M=D{\!\!\!\!/\,}[A]+\mu\gamma_0+m$ with 
$\mu$ the quark chemical potential, results in a complex-valued 
path integral measure $\mbox{det}M$ when
$\mu\not=0$. Since $\mu>0$ promotes baryon current flow in the 
positive $t$-direction, 
the fundamental reason for this {\em Sign Problem\/}
can be traced to the explicit
breaking of time reversal symmetry. 
Because the measure no longer has an interpretation as a
probability distribution, Monte-Carlo importance sampling, 
the mainstay of lattice
simulations, is completely ineffective in the thermodynamic limit.

It is instructive to ask what goes wrong when simulations are performed with a
measure $\mbox{det}M^\dagger M$ which is positive definite by construction, as
is the case for all practical fermion algorithms. It turns out that while $M$
describes a color-triplet quark $q\in\bf{3}$, $M^\dagger$ describes a {\em
conjugate quark\/} $q^c\in\bf{\bar3}$. The model's spectrum thus contains
gauge-singlet $qq^c$ states, indistinguishable from mesons at $\mu=0$, but 
carrying non-zero baryon number. As $\mu$ rises, baryonic matter
first appears in the ground state (i.e. $n_q>0$)
at an onset $\mu_o\sim{1\over2}m_\pi$, 
i.e. with an energy per quark comparable with the lightest baryon in the
spectrum, which is degenerate with the pion, rather than the physically
expected $\mu_o\sim{1\over3}m_{nucleon}$. 
Only calculations performed with the correct
measure $\mbox{det}^{N_f}M$ have cancellations among configurations,
due to the fluctuating phase of the determinant, which ensure that
$n_q$ vanishes for ${1\over2}m_\pi<\mu<{1\over3}m_{nucleon}$.

For Two Color QCD (QC$_2$D), ie. for gauge group SU(2), this bug is actually
a feature. Since $q$ and $\bar q$ live in equivalent representations of the
color group, hadron multiplets contain both $q\bar q$ mesons and $qq$ baryons.
It is correspondingly straightforward to show that the quark determinant 
is positive definite for even $N_f$ \cite{Hands:2000ei}. QC$_2$D is thus the
simplest model of dense strongly-interacting matter amenable to study with
orthodox lattice techniques. Additionally,
if there is a separation of scales $m_\pi\ll
m_\rho$ in the spectrum, then at low densities attention may be focussed on the 
Goldstone bosons of the system (both mesons and baryons) using chiral
perturbation theory ($\chi$PT) \cite{Kogut:2000ek}. The key result is that for
$\mu\geq\mu_o={1\over2}m_\pi$, a non-vanishing quark density $n_q>0$ develops,
along with a superfluid diquark condensate $\langle qq\rangle\not=0$. Just above
onset, the system is thus a textbook Bose Einstein Condensate (BEC) formed from 
tightly bound scalar diquarks.

Using the $\chi$PT prediction for $n_q(\mu)$ \cite{Kogut:2000ek}, it is simple
to develop the full equation of state, i.e. pressure $p$ and
energy density $\varepsilon_q$, at $T=0$ \cite{Hands:2006ve}:
\begin{eqnarray}
n_q&=&8N_ff_\pi^2\mu\Bigl(1-{\mu_o^4\over\mu^4}\Bigr);\nonumber\\
p=\textstyle\int_{\mu_o}^\mu n_qd\mu &=& 4N_ff_\pi^2
\Bigl(\mu^2+{\mu_o^4\over\mu^2}-2\mu_o^2\Bigl);\label{eq:model}\\
\varepsilon_q=-p+\mu n_q &=& 4N_ff_\pi^2\Bigl(\mu^2-3{\mu_o^4\over\mu^2}+
2\mu_o^2\Bigr);\nonumber\\
\langle qq\rangle &\propto& \sqrt{1-{\mu_o^4\over\mu^4}}.\nonumber
\end{eqnarray}
Here, $f_\pi$ is a parameter of the model. Contrast this with another
paradigm for cold dense matter, namely a degenerate system of weakly-interacting
massless
quarks populating a Fermi sphere up to some maximum momentum $k_F\approx\mu$:
\begin{equation}
n_q={{N_fN_c}\over{3\pi^2}}\mu^3;\;\;\;
\varepsilon_q=3p={{N_fN_c}\over{4\pi^2}}\mu^4.
\label{eq:SB}
\end{equation}
Superfluidity in this scenario arises from the condensation of quark Cooper
pairs within a layer of thickness $\Delta$ centred on the Fermi surface, so that
$\langle qq\rangle\propto\Delta\mu^2$.

\begin{figure}
\begin{center}
%\resizebox{0.75\textwidth}{!}{%
\includegraphics*[width=\colw]{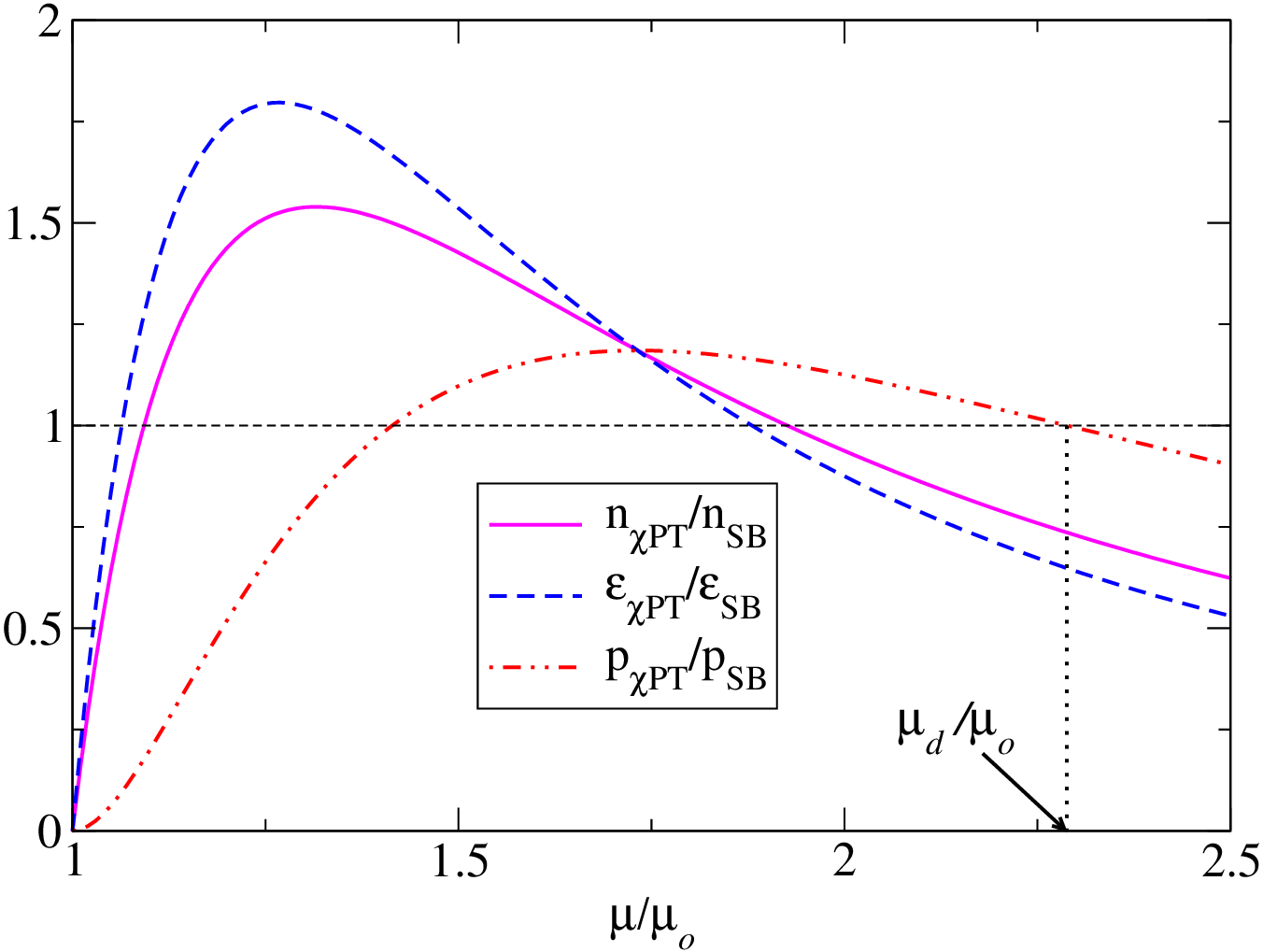}
%}
\label{fig:model}       % Give a unique label
\caption{Model equation of state for $f_\pi^2=N_c/6\pi^2$}
\end{center}
\end{figure}
Fig.~\ref{fig:model} plots $n_q$, $p$ and $\varepsilon_q$ from
(\ref{eq:model}), each divided by the 
free field results (\ref{eq:SB}), as functions of $\mu$. On equating pressures,
this naive model, which ignores all non-Goldstone and gluonic degrees of
freedom, predicts a first order deconfining transition from BEC
to ``quark matter'' at $\mu_d\approx2.3\mu_o$ with the choice 
$f_\pi=N_c/6\pi^2$.

\section{Simulation}
To test whether this prediction holds in a more systematic calculation we have
performed simulations of SU(2) lattice gauge theory with $N_f=2$ 
Wilson fermions with $\mu\not=0$ \cite{Hands:2006ve}. 
The Wilson formulation is not obviously a 
stupid choice: Wilson fermions retain a conserved baryon charge; any problems
with chiral symmetry should dominate in the low-$k$ region of the quark
dispersion curve, which lies at the bottom of the Fermi Sea and is hence inert;
moreover, studies with free fermions show that saturation artifacts due to 
the complete filling of the first Brillouin zone actually set in at higher
values of $\mu$ than is the case for staggered \cite{Bietenholz:1998rj}.
Most importantly, the eigenvalue spectrum of the Wilson Dirac operator
has the same symmetries as that of continuum QC$_2$D. As shown in
\cite{Hands:2006ve}, this fact permits an exact ergodic hybrid Monte Carlo 
algorithm for $N_f=2$, with no requirement to take a fourth root,
which may be problematic for $\mu\not=0$ \cite{Golterman:2006rw}.
The only novelty of our simulation is the inclusion of a diquark source term
\begin{equation}
jqq\equiv j\kappa(-\bar\psi_1(x)C\gamma_5\tau_2\bar\psi_2^{tr}(x)+
\psi_2^{tr}(x)C\gamma_5\tau_2\psi_1(x))
\end{equation}
in the dynamics, where subscripts label flavor and the Pauli matrix acts on
color. As well as making the algorithm ergodic, setting $j\not=0$ mitigates 
the effect of IR fluctuations due to Goldstone modes in any superfluid
phase, and of course enables direct estimation of the $\langle qq\rangle$
condensate.

Our initial study has been performed on an $8^3\times16$ lattice using 
a standard
Wilson gauge action, with parameters $\beta=1.7$, $\kappa=0.178$, and 
$j=0.04$ (with a few points taken at $j=0.02$, 0.06). Studies of the static 
quark potential 
and the hadron spectrum at $\mu=0$ yield $a=0.220$fm, $m_\pi a=0.79(1)$, and
$m_\pi/m_\rho=0.80(1)$.~\footnote{this corrects the value erroneously given in
\cite{Hands:2006ve}} We thus expect the onset of baryonic matter at
$\mu_oa\approx0.4$. Thermodynamic observables are calculated as follows: quark
density is given by a local operator via
\begin{equation}
n_q=-{{\partial\ln{\cal Z}}\over{\partial\mu}}.
\end{equation}
As a component of a conserved current, it is immune from quantum corrections,
but may be affected by artifacts due to $a>0$, $V<\infty$. We therefore prefer
to quote our results in terms of $n_q/n_{SB}^{\rm latt}$, where $n_{SB}^{\rm
latt}(\mu)$ 
is evaluated for free massless quarks on the same lattice. The pressure
follows from an integral formula
\begin{equation}
{p\over p_{SB}}=\int_{\mu_o}^\mu{{n_{SB}^{\rm cont}}\over{n_{SB}^{\rm latt}}}
n_qd\mu \biggl/ \int_{\mu_o}^\mu n_{SB}^{\rm cont}d\mu.
\end{equation}
Note that although $p$ is calculated purely in terms of quark observables, 
it is in principle the pressure of the system as a whole, although both
continuum and thermodynamic limits must eventually be taken.
Finally, quark energy density is also estimated by a local operator
\begin{equation}
\varepsilon_q=\kappa\Bigl\langle
\bar\psi_x(\gamma_0-1)e^\mu U_{0x}\psi_{x+\hat 0}
-\bar\psi_x(\gamma_0+1)e^{-\mu}U_{0x-\hat 0}^\dagger\psi_{x-\hat 0}
\Bigr\rangle;
\end{equation}
this requires both subtraction of the $\mu=0$ vacuum contribution, 
and a $\mu$-independent but as yet
unknown multiplicative renormalisation. In what follows, therefore, the
shape of the curve is in principle correct, but the overall scale still
undetermined.

\begin{figure}
\begin{center}
\includegraphics*[width=\colw]{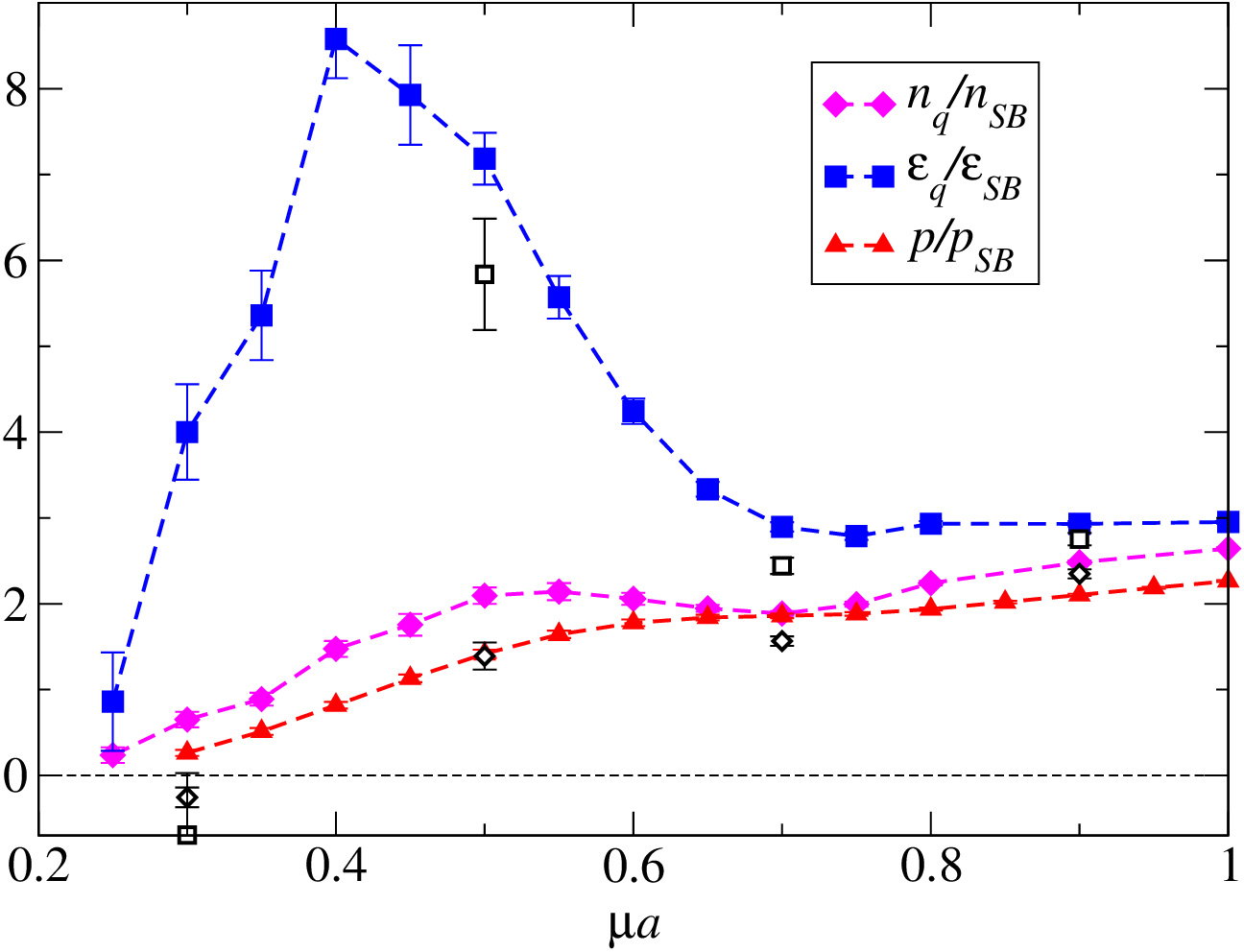}
\label{fig:eneps}       % Give a unique label
\caption{Lattice equation of state for $j=0.04$ (open symbols give $j\to0$
extrapolation}
\end{center}
\end{figure}
Fig.~\ref{fig:eneps} summarises our results. Both $n_q$ and $p$ start to rise 
from zero at $\mu a\approx0.3$, 
although a careful $j\to0$ extrapolation will be
needed to pinpoint the onset with any precision. By $\mu a\approx0.5$ both
quantities scale with $\mu$ in general accordance with 
free-field predictions, but with approximately twice the expected value. One
explanation of this mismatch is that the system has formed a Fermi sphere with
$\mu=E_F<k_F\propto n_q^{1\over3}$, which could be attributed to a negative
binding contribution to $E$ from interactions. The quark energy density, 
by contrast,
increases more slowly than free-field expectations up to $\mu a\approx0.65$,
whereupon free-field scaling sets in rather abruptly. Another intriguing result
\cite{Hands:2006ve} is that for $0.4\leq\mu a<1.0$ the gluon energy density
$\varepsilon_g$ (identically zero in free-field theory) scales to quite high
precision as $\mu^4$, the only physically sensible
possibility once $\mu/T\gg1$. Note that $\varepsilon_g>0$ entirely as a result 
of interactions with the background quark density, 
since this is the only means by which
$\mu$-dependence can arise.

\begin{figure}
\begin{center}
\includegraphics*[width=\colw]{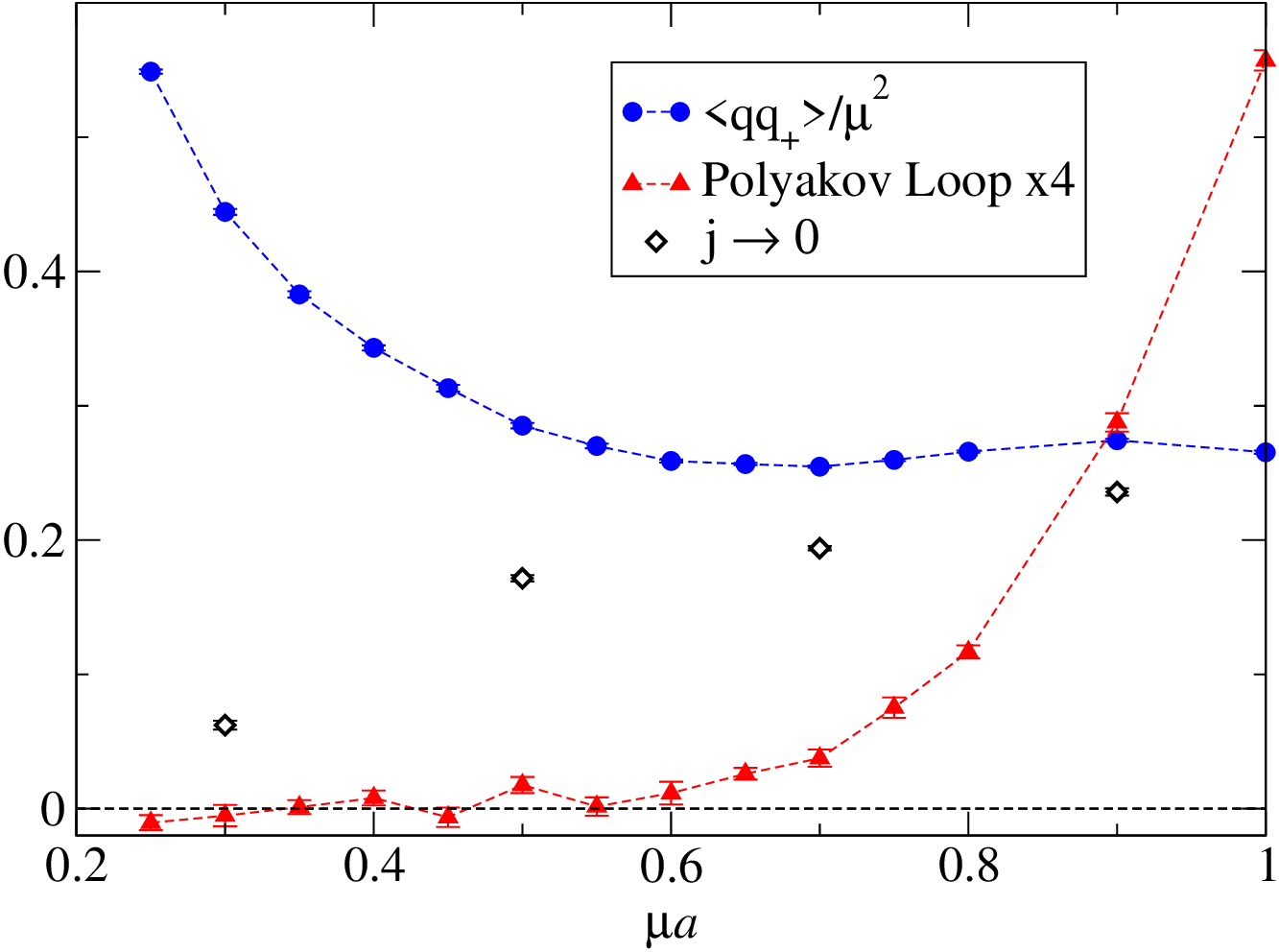}
\label{fig:orderps}       % Give a unique label
\caption{Order parameters $\langle qq\rangle$ and $L$ vs. $\mu$}
\end{center}
\end{figure}
To elucidate what's happening, Fig.~3 plots both the 
superfluid order parameter $\langle qq\rangle$ divided by $\mu^2$, and the
Polyakov line $L$. For $\mu a\geq0.5$ it is clear the system is in a superfluid
phase, but what is remarkable is that at $\mu a\approx0.6$ there is a sudden 
transition to a regime where $\langle qq\rangle\propto\mu^2$, as
expected for BCS pairing at a Fermi surface. At roughly the same point $L$ 
rises from zero; although for theories with fundamental matter $L$ is not
strictly an order parameter, this is suggestive that at
$\mu a\approx0.65$ there is a {\em deconfining\/} transition, beyond which the
effective degrees of freedom are best thought of as quarks (or even
{\em quasiquarks\/}), and not
the scalar diquarks of $\chi$PT.

\section{Discussion}

Our initial study of thermodynamic quantities, and of the
properties of the ground state, strongly suggests 
that QC$_2$D at low temperature
has at least two transitions as chemical potential $\mu$ is raised. The first is
between the vacuum and a phase of Bose-condensed tightly-bound diquarks; the
second, a relativistic analogue of the BEC/BCS crossover currently discussed in
both strongly-correlated electron and cold atom systems, is a deconfining
transition to a system of degenerate quarks, the Fermi surface being mildly
disrupted by a Cooper pair condensate. Although QC$_2$D clearly models
nuclear matter unrealistically, 
its description of quark
matter may well prove to have much in common with that of QCD. 
We are currently extending our study to
the hadron spectrum, and to finer lattice spacings to check that this conclusion
is not due to lattice artifacts. Interesting results obtained from a study of
the gluon propagator on the current system will be discussed elsewhere
\cite{Jon-Ivar,Hands:2006ve}

\begin{figure}
\begin{center}
\includegraphics*[width=\colw]{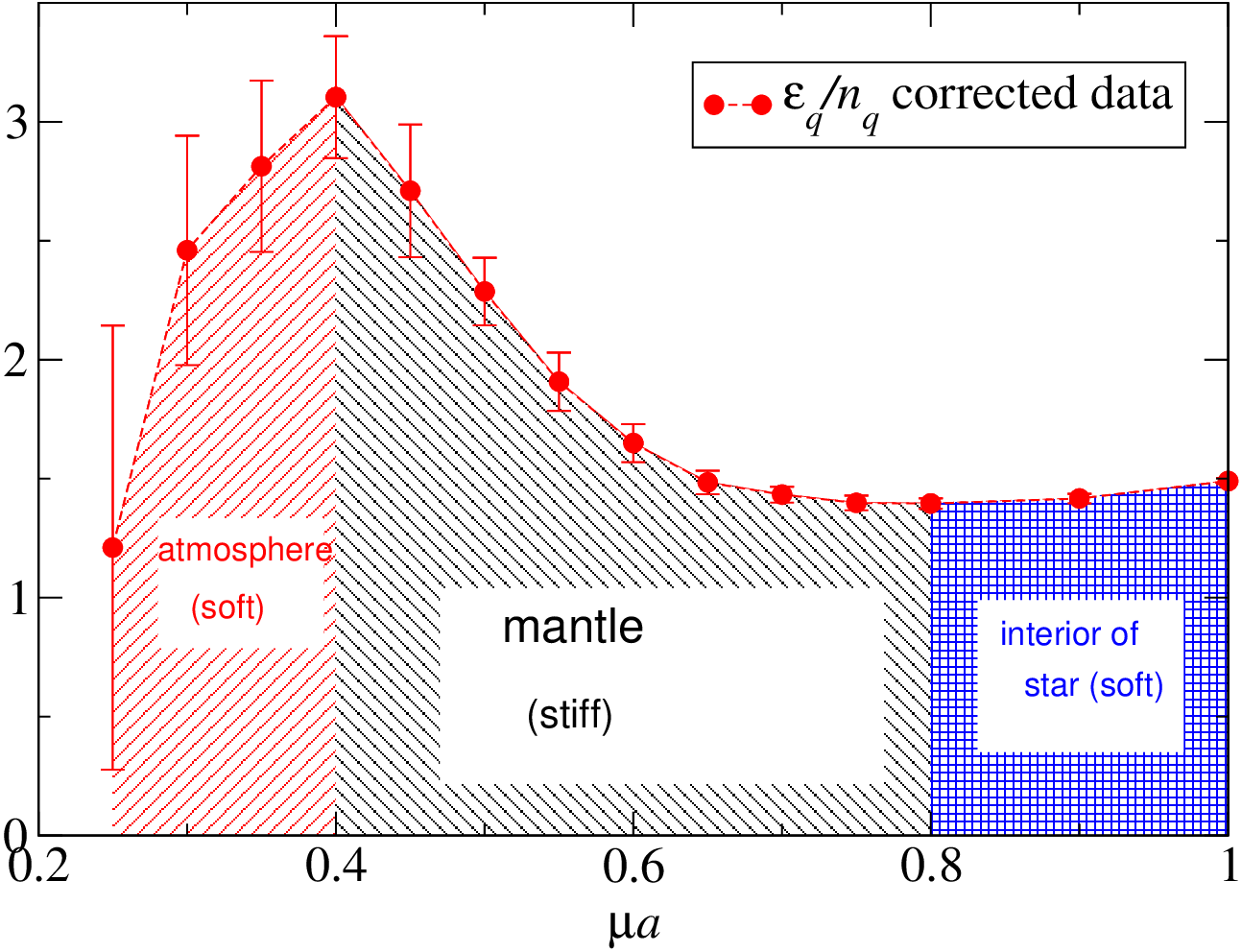}
\label{fig:star}       % Give a unique label
\caption{Energy per quark $\varepsilon_q/n_q$ vs. $\mu$}
\end{center}
\end{figure}
Meanwhile it is hard to resist the temptation to speculate on what a Two Color
Star might look like. Fig.~\ref{fig:star} plots the energy per quark
$\varepsilon_q/n_q$ versus $\mu$ using the data of Fig.~\ref{fig:eneps}. The
most striking feature of this plot is the pronounced minimum at 
$\mu a\approx0.8$,
which is both robust (since it occurs even if corrections for $a>0$, $V<\infty$
are left out), and unexpected (since it does not occur for the model EoS 
of Fig.~\ref{fig:model}). We infer that any large object
assembled from a fixed number of QC$_2$D quarks,
such as a star, will have the bulk of its interior in the
neighbourhood of this minimum, which as Fig.~3 shows, means that the object
would
in effect be a quark star formed from deconfined matter. 
Somewhat speculatively, we
have labelled the different regions of the $\mu$-axis with the corresponding 
layers of the star, although a quantitative solution for the radial profile must
await correctly-normalised calculations of the energy densities $\varepsilon_q$
and $\varepsilon_g$.

\end{document}